\documentclass[11pt, letterpaper]{article}
\usepackage{placeins}

\usepackage[margin=1.5cm]{geometry}
\usepackage{titlesec}
\usepackage{tabu}
\usepackage{enumitem}
\usepackage{amssymb}
\usepackage[square, numbers, sort&compress]{natbib}
\usepackage{xcolor}
\newlist{selectlist}{itemize}{2}
\setlist[selectlist]{label=$\square$,leftmargin=*,noitemsep,topsep=0pt}
\usepackage{lscape}
\usepackage{float}

\usepackage{lmodern}

\usepackage{graphicx}
\usepackage{subcaption}

\usepackage{todonotes}
\usepackage{siunitx}
\usepackage{appendix}

\usepackage{hyperref}
\hypersetup{
    colorlinks=true,
    linkcolor=blue,
    filecolor=magenta,      
    urlcolor=blue,
}
 
\titleformat{\section}[block]{\hspace{1em}\bfseries}{\thesection.}{0.5em}{} 
\titleformat{\subsection}[block]{\hspace{1em}}{\thesubsection}{0.5em}{}

\begin{document}


\setlength{\parindent}{0pt}
\setlength{\parskip}{10pt}

\textbf{Article title}

Active Cooling Device: A Flexible, Lab-Scale Experimental Unit to Develop Spatio-Temporal Temperature Control Strategies

\textbf{Authors}

Victor Oliveira Ferreira$^{1,3}$, Wiebke Mainville$^{1,3}$, Vincent Raymond$^2$, Jean-Michel Lamarre$^2$, Antoine Hamel$^2$, Mikael Vaillant$^{1,3}$, Moncef Chioua$^{3}$, Bruno Blais$^{1,3,*}$

\textbf{Affiliations}

\begin{enumerate}
    \item Chemical engineering High-performance Analysis, Optimization and Simulation (CHAOS), Department of Chemical Engineering, Polytechnique Montr\'eal, PO Box 6079, Stn Centre-Ville, Montr\'eal, QC, Canada, H3C 3A7.

    \item National Research Council Canada, Boucherville, Québec, Canada.

    \item Department of Chemical Engineering, Polytechnique Montr\'eal, PO Box 6079, Stn Centre-Ville, Montr\'eal, QC, Canada, H3C 3A7.
\end{enumerate}

\textbf{Corresponding author’s email address}\\ {*}bruno.blais@polymtl.ca

\textbf{Abstract}\\

We present an experimental unit that realizes the ``multi-input, multi-output manifold'' thermal management technology proposed by \citet{Lamarre_2023}. The proposed setup can be used for experiments aimed at controlling spatiotemporal temperature distribution. Temperature control is achieved by impinging coolant fluid jets, leveraging a manifold of channels targeted to the surface. The direction of the fluid is controlled by shifting the role of channels between inputs, outputs, or closing them. Files associated with this work include Computer-Aided Design (CAD) STEP files, Gerber files to manufacture a Printed Circuit Board (PCB), and a Graphical User Interface (GUI) written in Python. We provide a step-by-step guide to assemble the experimental setup. We also provide instructions to interact with the setup through the GUI, which allows for real-time tracking of sample temperature and flow rates per flow control device. Additionally, we provide examples of usage of the setup, including system characterization with step response, Proportional-Integral-Derivative performance tracking, and disturbance rejection in a coupled system. Extending the application is accessible through the files provided in the open repository associated with this work. The active cooling device presents a safe, flexible, and complete design, allowing for lab-scale assessment of the performance of custom temperature control strategies using enclosed impinging jets.


\newpage
\textbf{Specifications table}\\
\vskip 0.2cm
\tabulinesep=1ex

\begin{tabu} to \linewidth {|X|X[2,l]|}
\hline  \textbf{Hardware name} & Active cooling experimental setup
  \\
  \hline \textbf{Subject area} & %
  \begin{itemize}[noitemsep, topsep=0pt]
  \item Manufacturing and process engineering
  \item Process control
  \item Thermal management
  \end{itemize}
  \\
  \hline \textbf{Hardware type} & %
  \begin{itemize}[noitemsep, topsep=0pt]
  \item Thermal control device
  \item Lab-scale twin
  \end{itemize}
  \\ 
\hline \textbf{Closest commercial analog} & No commercial analog is available.\\
\hline \textbf{Open source license} & BSD 3-Clause License \\
\hline \textbf{Cost of hardware} &
  Approximately 14000.00 USD, including only the price of off-the-shelf material. Costs do not include manufacturing and design.
  \\
\hline \textbf{Source file repository} & 
  \href{https://doi.org/10.5281/zenodo.15644038}{https://doi.org/10.5281/zenodo.15644038}
\\\hline
\end{tabu}

\newpage
\section{Hardware in context}

Thermal management is among the most important and sensitive operations in several industries. From asserting part quality upon manufacturing to optimizing the performance of electronic components and batteries, cooling and heating systems are key components in processes involving high thermal loads and chemical reactions. For example, in die casting or pressure molding, the mechanical properties of the manufactured parts are highly dependent on the cooling rate \cite{Wang_1995}. When this cooling rate is poorly controlled, parts can suffer uncontrolled shrinkage and might require additional heat treatment to achieve the desired properties. As most parts present irregular geometry and uneven mass distribution, accounting for the shape of the part while designing the mold's cooling system is challenging. One way of doing this is using conformal cooling channels, an effective yet inflexible, case-sensitive method that requires advanced manufacturing technology such as 3D printing \cite{Feng_2021}.

The electrification of vehicles presents another example underlining the importance of thermal management. Lithium-ion batteries (LIBs) are considered one of the most viable energy storage devices for electric vehicles due to their longevity and energy storage capacity \cite{Kim_2019}. However, their viability requires dense energy cells, implying high heat generation per cycle per cell. As widely reported, LIBs deteriorate more rapidly when subjected to high temperatures \cite{Kim_2019}. This deterioration increases the batteries' impedance, causing overheating. In contrast, low temperatures reduce LIBs' discharge capacity \cite{Kim_2019}. Hence, poor thermal management drastically decreases LIBs' performance, limiting the application of technologies that depend on high-density energy storage systems, such as electric vehicles.

For these reasons, developing an adaptive temperature control device is key for sensitive industries. In this context, \citet{Lamarre_2023} proposed a flexible active cooling system, consisting of a chamber with a manifold of inlet/outlet pipes applied to control the flow of a cooling and heating fluid. The active temperature control device is designed to have control, in time and space, of the temperature profile across a surface. This novel technology can be used, for instance, to impose higher cooling rates at higher temperature regions, ensure even temperature distribution on a surface regardless of the heat load, or even serve as a local heat source for parts with uneven shapes or regions where losses are higher.

\begin{figure}[H]
        \centering
        \includegraphics[width=0.8\linewidth]{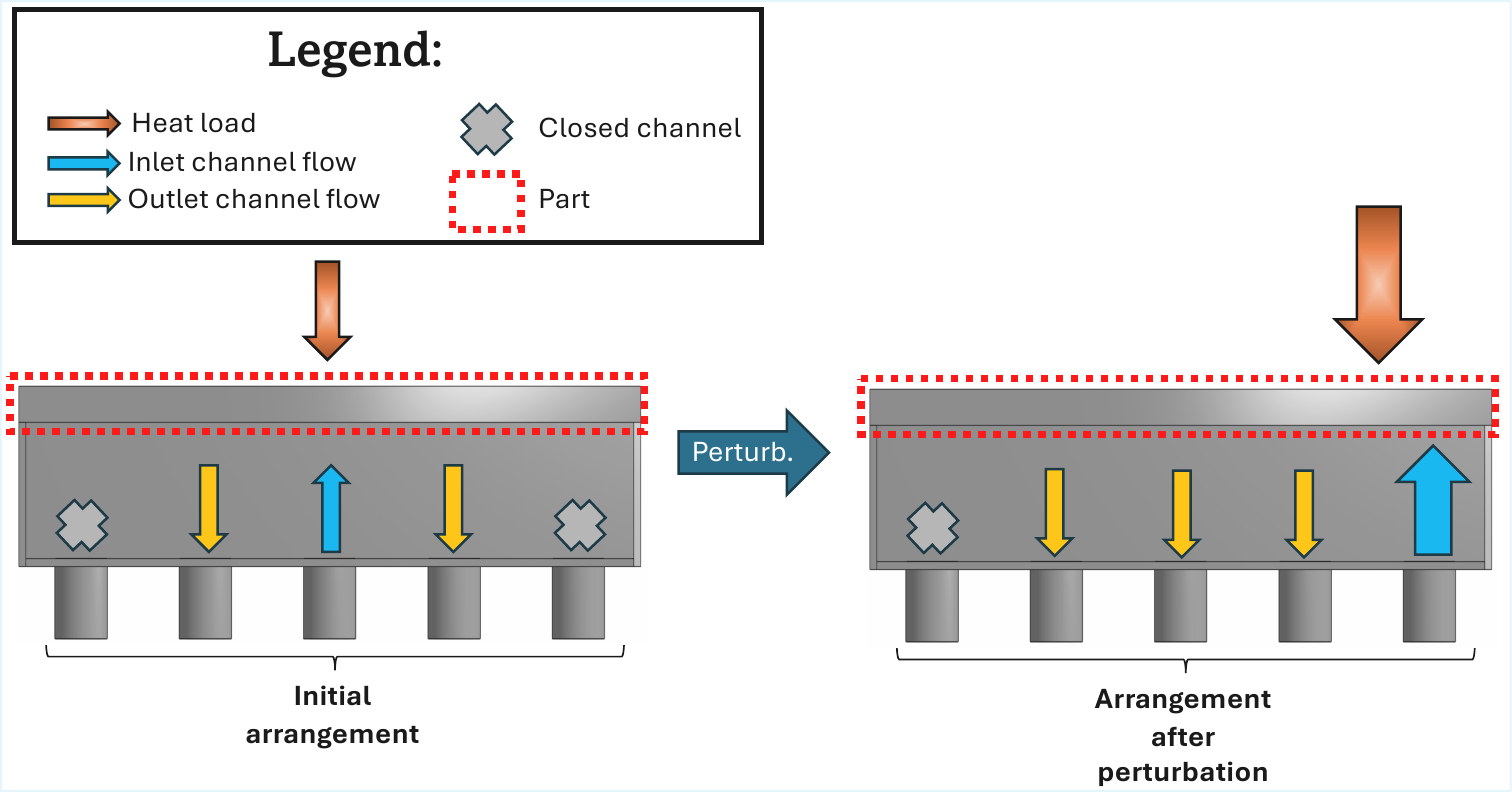}
        \caption{Illustration of the perturbation rejection principle of the active cooling device. On the left, the system is at steady-state using the initial arrangement. When the heat load is shifted to the right and increased in magnitude, the system reacts, shifting the arrangement and adjusting the inlet flow rate accordingly.}
        \label{fig:context_schematic}
\end{figure}

The principle behind the temperature control is illustrated in Figure \ref{fig:context_schematic}. In the figure, the part we aim to control the temperature of is highlighted by the red dotted rectangle. The part is exposed to a given heat load represented by the orange arrows pointing down to the plate. The temperature of the part is manipulated by controlling the fluid flow within the chamber below the part, represented by the arrows and ``x'' symbols within it. In the illustrated device, each of the five channels can be an inlet, an outlet, or closed.

In the process in Fig. \ref{fig:context_schematic}, the target temperature profile is achieved by injecting fluid by the central channel (blue arrow pointing up to the plate), opening its neighbors to be outlets (yellow arrows pointing down from the plate), and closing the channels at the extremities (represented by ``x'' symbols). After a perturbation, the device should react by changing the inlet/outlet/closed channel arrangement and the inlet flow rate (represented by the size of the arrow) in order to maintain the target temperature profile.

A device using a similar principle has been reported by \citet{Hopmann_2020} and applied for polymer pressure molding. Nevertheless, their device differs from the one proposed by \citet{Lamarre_2023} on the heat transfer mechanism, as the former uses an element to heat the injected polymer by conduction while cooling is done using gas expansion, whereas the latter applies forced convection by jet impingement. The enclosed impinging jets can interact more or less depending on the operational parameters applied (such as fluid temperature, the jet flow rate, and the ratio between the diameter of the jet and its distance to the surface) \cite{Wae-Hayee_2015, Barbosa_2023}, which can imply non-trivial interaction between controllers actuating in different regions.

While the novel active cooling device is promising, it is not clear what the best control strategy for each targeted application. Additionally, testing strategies in production can be resource-intensive and risky. As such, its application is limited due to the lack of prior testing.

In this work, we present a lab-scale version of the active cooling device, which can be safely used for testing different operational conditions and control strategies under varying heat loads. Among the provided data, we included the schematics of a custom Printed Circuit Board (PCB), which integrates the electronic components of the hardware. We also provide the code used to control the setup, including the entire backend and a Graphical User Interface (GUI) coded in Python with PySide6, which is available under BSD 3-Clause license. Lastly, we present some results of preliminary experiments, which are used to illustrate potential applications of the technology.

\section{Hardware description}
\begin{figure}[H]
        \centering
        \includegraphics[width=0.8\linewidth]{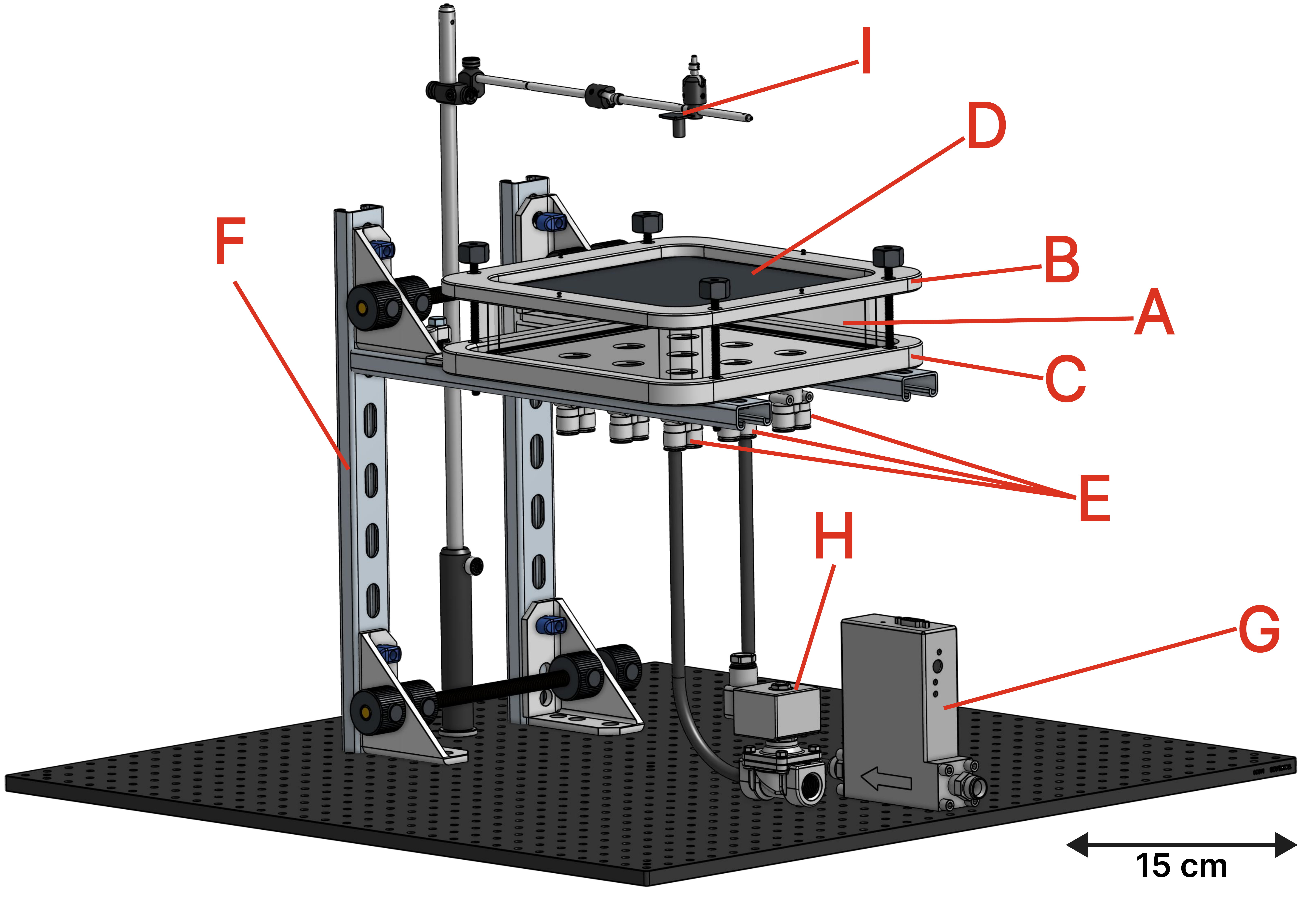}
        \caption{Computer-aided design representation of the experimental setup.}
        \label{fig:cad_iso}
\end{figure}

The active cooling device uses enclosed impinging jets to control the temperature profile of a metallic plate in both time and space. In this section, we describe the device's parts and their respective roles. Figure \ref{fig:cad_iso} shows an isometric illustration of the experimental setup. As shown in the figure, the device is divided into three units, which we call the heat source, the chamber, and the control hardware.

\begin{itemize}
    \item \textit{Heat source} - The heat source can vary with the targeted application. In the present work, we use a commercially available heat gun (available on McMaster-Carr under the serial number 3433K56) with air temperature control. The heat gun was chosen for its portability, ability to provide a consistent localized heat load, and its non-blocking path for the infrared camera line-of-sight. The setup can be adapted to a wide range of heat sources, as long as a surface temperature distribution measurement can be obtained.

    \item \textit{Chamber} - The enclosed chamber (A) is sealed by compressing it between the upper part (B) and the lower (C) using threaded rods. The metal plate (D) is positioned between the upper part (B) and the enclosed chamber (A), while rubber o-rings on each side of the chamber (A) are used to prevent leakage. Y-shaped tube fittings (wye, E) are connected to the bottom of the chamber. Each of those wyes has the same configuration:  one side connects the chamber to the Mass Flow Controller (MFC, G), connected to the compressed air line, while the other end is connected to a solenoid valve (H), which opens to ambient air. Figure \ref{fig:cad_iso} shows only one MFC and one solenoid valve, but each wye is connected to its own pair. These connections enable various jet arrangements, as described in the following subsection. Finally, the enclosed chamber is supported by the chamber mount (F), mounted to an optical table.

    \item \textit{Control hardware} - Comprises the sensors (MFC monitor and IR camera), actuators (MFC control valves and solenoid valves), the custom Printed Circuit Board (PCB), and a Raspberry Pi (RPi). The RPi communicates with the hardware parts, collecting measurements and acting on the jets according to the implemented algorithm. Details on components are provided in the following dedicated subsections.
\end{itemize}

Following, we explain the operation of the equipment and the function of each unit. Furthermore, details about the rationale behind the design choices and their associated limitations are provided.

\subsection{Jet flow rate and arrangements}

The temperature control is done by manipulating two variables, namely jet arrangement (or arrangement, for brevity) and flow rate. Each of the orifices at the bottom of the chamber can act as air outlets, jet inlets, or be shut. The arrangement of the inlet/outlet/shut orifices determines the distribution of the air flow within the chamber, which is used to manipulate the temperature distribution of the top plate (D). Additionally, the flow rate of the inlet jets can be modulated.

\begin{figure}[H]
        \centering
    \begin{subfigure}[b]{0.45\linewidth}
        \centering
        \includegraphics[width=\textwidth]{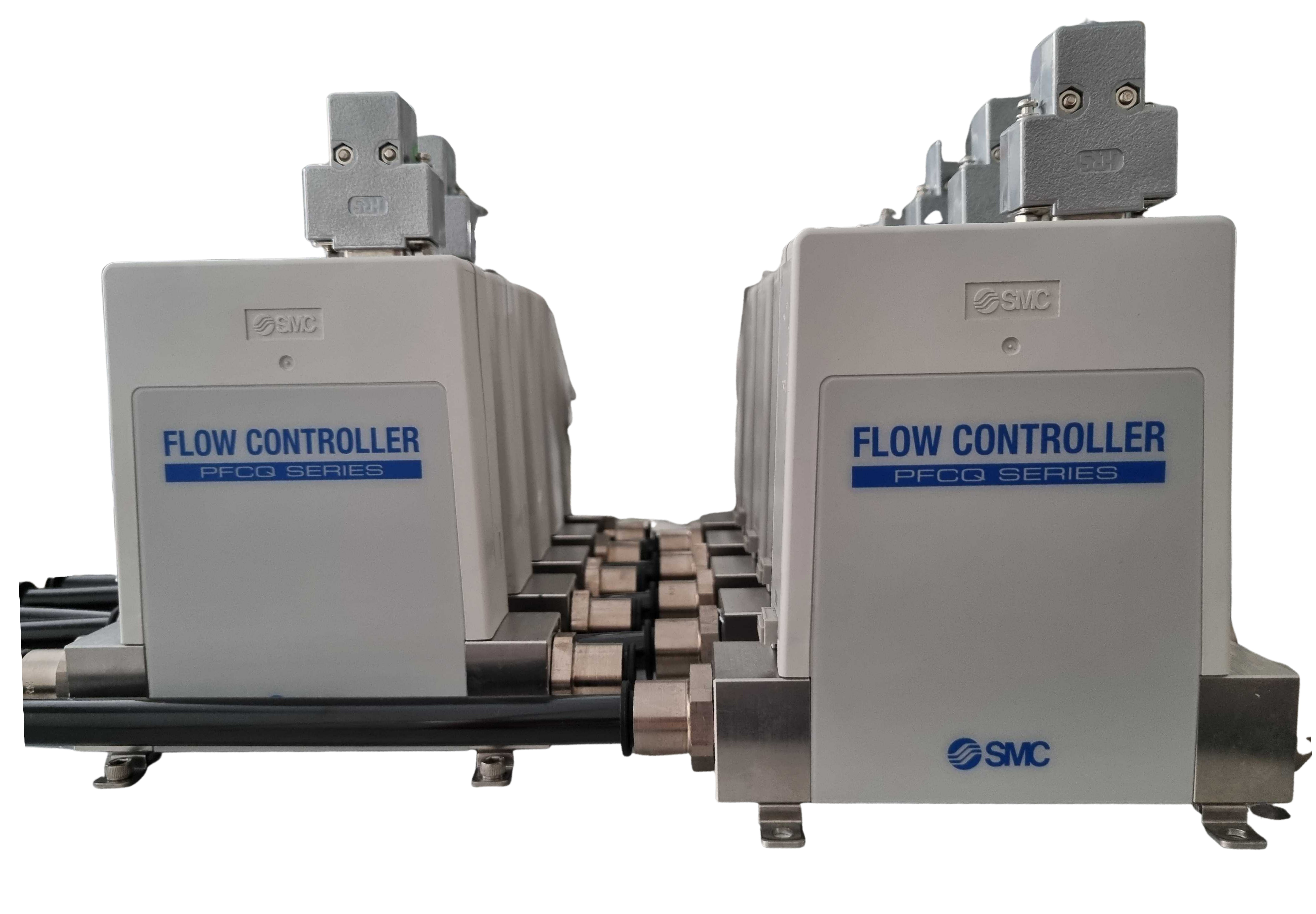}
        \caption{MFC.}
        \label{fig:mfc}
    \end{subfigure}
    \begin{subfigure}[b]{0.30\linewidth}
        \centering
        \includegraphics[width=\textwidth]{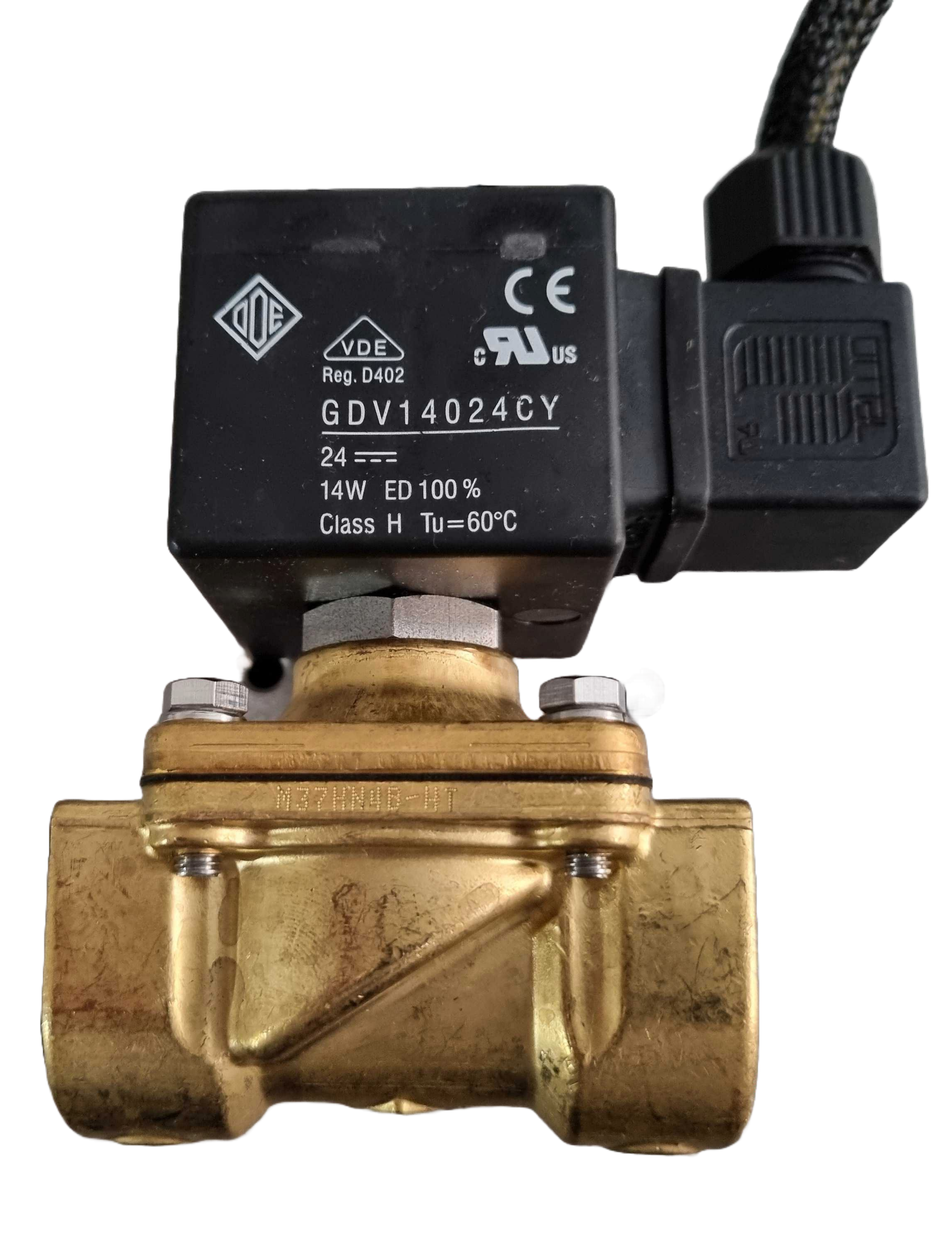}
        \caption{Solenoid valve.}
        \label{fig:solenoid}
    \end{subfigure}
        \caption{Picture of the flow control components used in the active cooling device.}
        \label{fig:mfc-solenoid}
\end{figure}

\begin{figure}[H]
        \centering
        \includegraphics[width=0.8\linewidth]{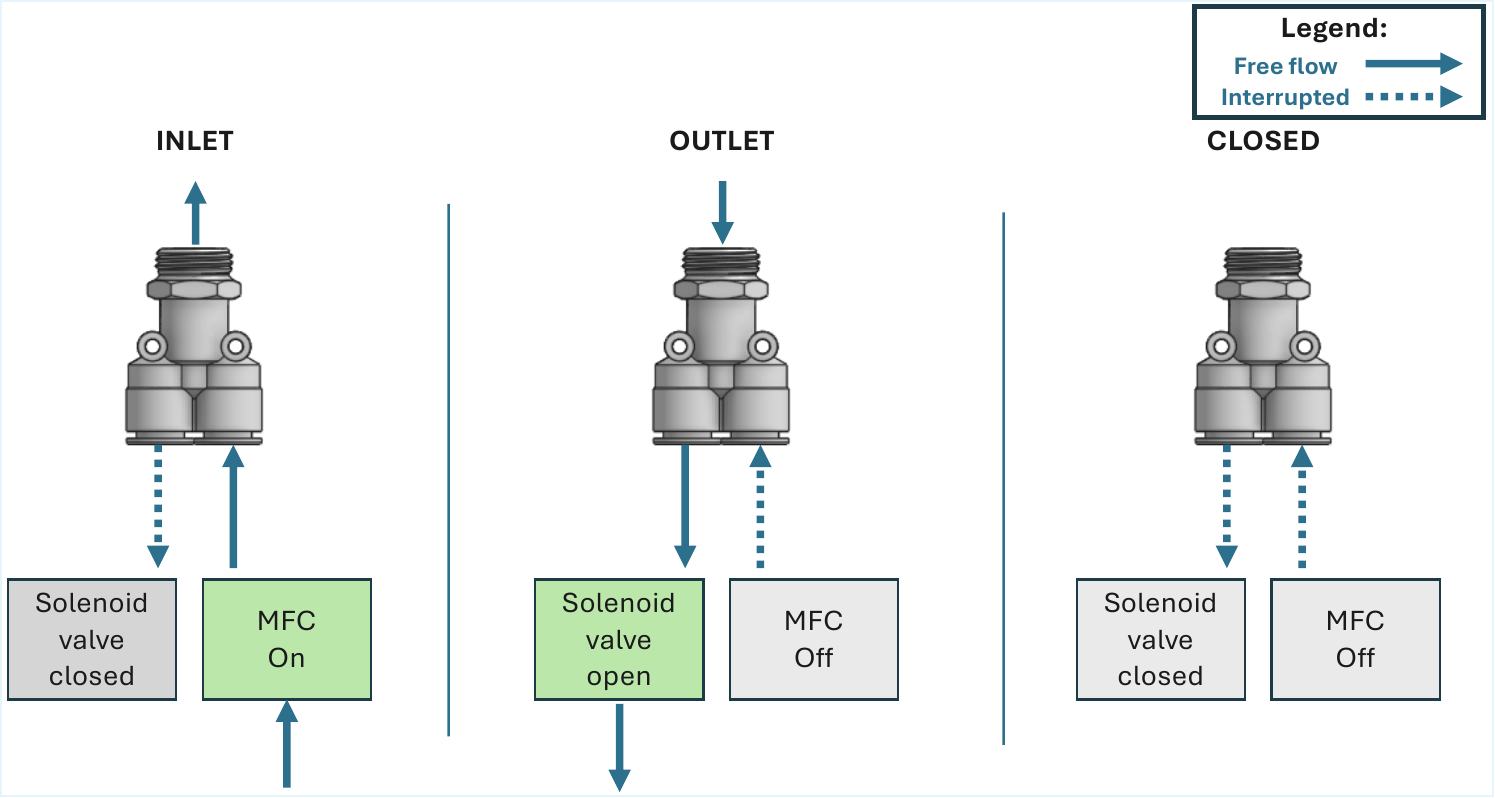}
        \caption{Schematic diagram depicting the jet arrangement manipulation logic.}
        \label{fig:arrangement-logic}
\end{figure}

The arrangement is controlled by the MFCs (Figure \ref{fig:mfc}) and the solenoid valves (Figure \ref{fig:solenoid}), concomitantly, as shown in Figure \ref{fig:arrangement-logic}. As previously noted, a pair of MFC-solenoid valve is connected to a wye flow splitter, which is connected to the orifice of the channel. When the orifice is in a closed state, the solenoid valve controlling the outlet flow is shut, and the MFC flow rate is set to zero. When the orifice is set as an outlet, the solenoid valve downstream of the chamber is open, allowing the passage of air. Lastly, when the orifice is in an inlet state, the solenoid valve is closed, and the MFC controls the inlet flow rate. The equipment operates with at least one orifice as an outlet to prevent pressurization.

The inlet air jet flow rates are controlled by mass flow controllers (MFCs). In this work, we used MFCs model PFCQ531-04-A1C-S from SMC, as depicted in Figure \ref{fig:mfc}. When operating within the nominal pressure range, MFCs independently measure and control the gas flow rate with a fast settling time ($\le$0.5s at 300 kPa) and a large flow rate modulation range (9 to 300 L/min). The RPi then sends commands and receives actual flow rate data for each MFC through analog signals.

\subsection{Temperature measurement}

\begin{figure}[H]
        \centering
        \includegraphics[width=0.4\linewidth]{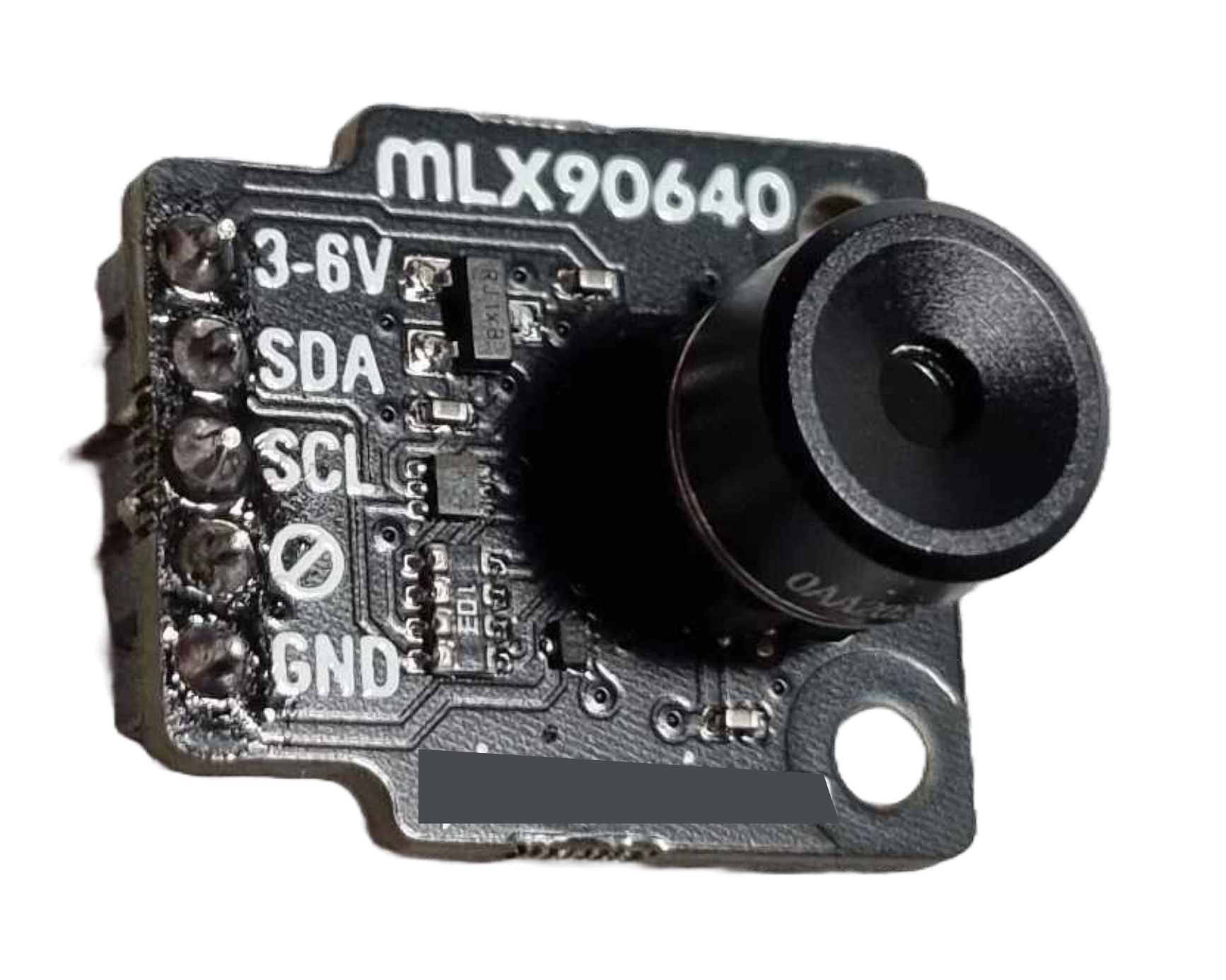}
        \caption{Module infra-red camera model MLX90640.}
        \label{fig:ircam}
\end{figure}

We use an MLX90640 infrared camera, depicted in Figure \ref{fig:ircam}, to measure the temperature across the metal plate's surface. The camera's resolution is 32x24, and it is available with field of view angles of 110\textdegree (MLX90640BAA) or 55\textdegree (MLX90640BAB). It communicates with the RPi through the I2C protocol, and the latter stores the temperature information as a vector. The temperature information is used for real-time control experiments.

\subsection{Printed Circuit Board}

\begin{figure}[H]
        \centering
        \includegraphics[width=0.8\linewidth]{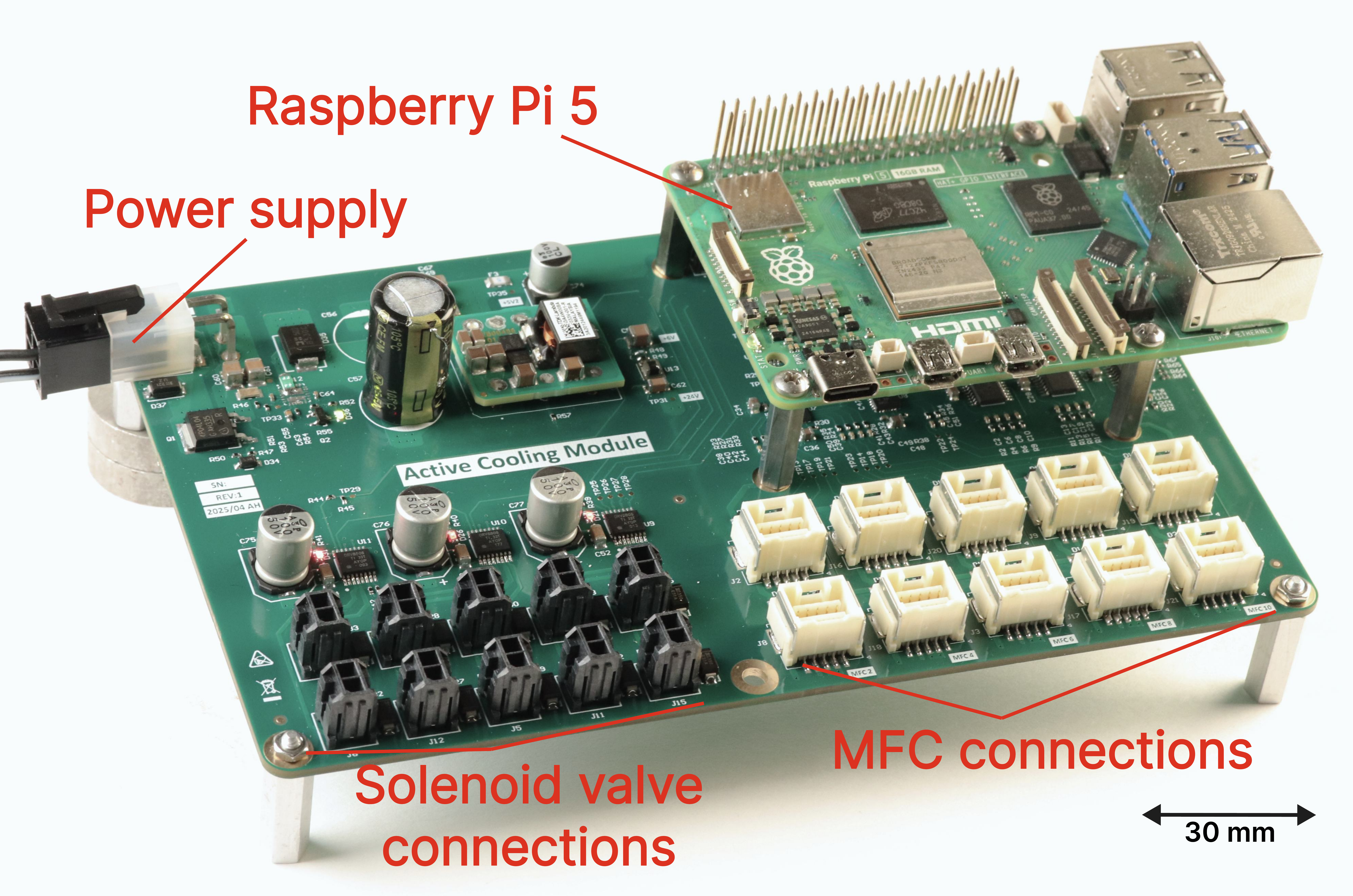}
        \caption{Picture of the custom PCB with mounted Raspberry.}
        \label{fig:pcb_picture}
\end{figure}

We use a custom Printed Circuit Board (PCB) to connect the electronic components to the sensors, valves, controllers, and the RPi. The custom PCB, depicted in Figure \ref{fig:pcb_picture}, was designed to feed the entire system using a single 24 \si{\volt} power supply, distributing the correct voltage for each component. The main chips include two Digital-Analog converters (DAC, model DAC7578SRGER), two Analog-Digital converters (ADC, model TLA2528IRTER), and three motor drivers for solenoid valves (model DRV8806). DAC and ADC modules communicate with the RPi through the I2C protocol, while the motor driver uses Serial-Peripheral Interface (SPI). A custom Python firmware was developed for communication with the three chips. The schematics, the code, and all information necessary to reproduce the module are fully available in the Zenodo repository \cite{ferreira_active_2025} associated with this publication.

\subsection{Graphical user interface (GUI)}

A GUI in Python was create and made availablesing the PySide6 library to facilitate interaction with the experimental setup. PySide6 was chosen because it is a reliable, high-level, lightweight interface between Python code and the Qt 6 library. Details about the interface are provided in Section \ref{sec:operation}.

\section{Design files summary}
All files described in this section are available on the Zenodo repository associated to this publication (DOI: 10.5281/zenodo.15644038) under Creative Commons Attribution 4.0 International.
\vskip 0.1cm
\tabulinesep=1ex
\begin{tabu} to \linewidth {|X|X|} 
\hline
\textbf{Design filename} & \textbf{File type} \\\hline
Active\_cooling\_CAD\_files.zip & CAD STEP files, including assembly of parts forming the setup.  \\\hline
Bill\_of\_materials.xlsx and Bill\_of\_materials.csv & Bill of materials in Excel and commas-separated values format.  \\\hline
PCB\_Design\_files.zip & PCB design files, including Gerber, footprint, descriptive Bill of Materials, and all other necessary files to build and assemble the PCB. \\\hline
\end{tabu}

\vskip 0.3cm
\noindent
The CAD step files in the repository can be imported in any STEP compatible CAD design software to recreate the design in Fig. \ref{fig:cad_iso}. We also included the Bill of Materials, which includes the price (at the time of construction) of each of the main parts used in the assembly. Finally, we also included the PCB design and construction files in the repository, which can be used to reproduce our assembly in Fig. \ref{fig:pcb_picture} in its totality. 

\section{Bill of materials summary}

The bill of materials is available in Appendix \ref{sec:bom} and on the repository online \cite{ferreira_active_2025}.

\section{Build instructions}

Most of the setup is composed of off-the-shelf components; therefore, assembling the experimental setup involves connecting the parts according to the design files in the Zenodo repository \cite{ferreira_active_2025} associated with this publication. The following section explains the process of assembling the setup.

\paragraph{Chamber mount:} The assembly starts by attaching the chamber mount to the optical table. The base is screwed to the optical table and the strut channel. After this, we screw together the vertical struts and \SI{90}{\degree} strut channel brackets. Then, we screw the horizontal struts to the strut channel brackets, which hold the chamber. Lastly, we use threaded rods to fix the space between vertical struts and prevent rotation.

\begin{figure}[H]
        \centering
        \includegraphics[width=0.8\linewidth]{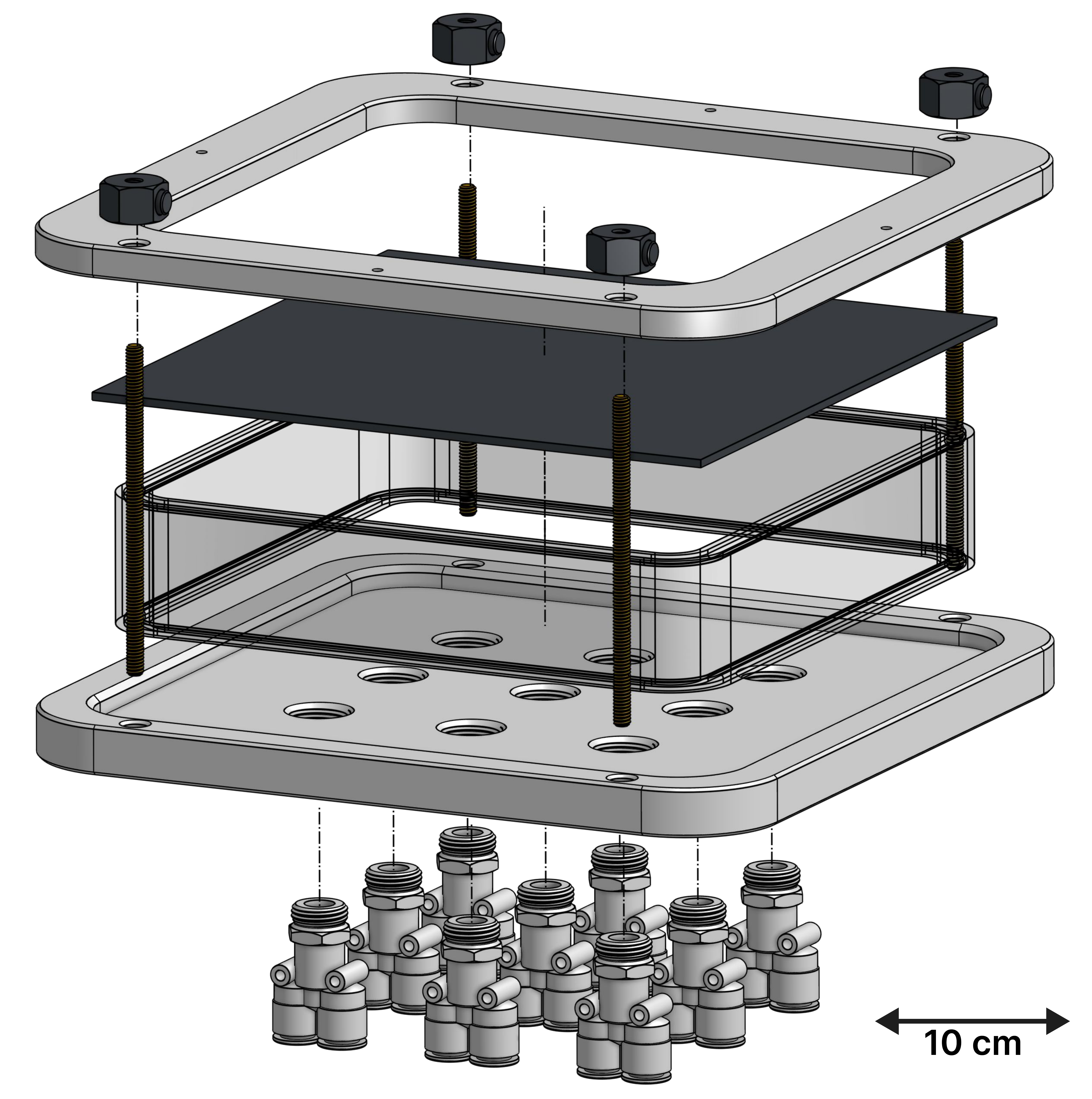}
        \caption{Exploded view of the chamber.}
        \label{fig:cad_exploded}
\end{figure}

\paragraph{Chamber:} We provide an exploded view of the chamber in Figure \ref{fig:cad_exploded}. The chamber is composed of the walls, the metallic plate sample, the top component (``hat''), and the base. First, we position the base on the strut. Then, we insert an o-ring into the grooves on either side of the acrylic chamber and place the chamber on top of the base. After this, we place the metallic plate on top of the chamber, followed by the hat. The complete component stack is held together and fixed to the mount with vertical threaded rods as shown in Figure \ref{fig:cad_exploded}.

\paragraph{Connection between compressed air line, MFCs, solenoid valve, and chamber:} Regardless of the compressed air source, the MFCs in this work require an upstream pressure between 50 and 800 \si{\kilo \pascal}, with a differential pressure range of 50 and 500 \si{\kilo \pascal}. To fulfill both pressure, differential pressure, and flow rate range requirements, we use 12 \si{\milli \meter} diameter flexible tubes to connect the MFCs to a compressed air line, supplying air at constant 450 \si{\kilo \pascal}. The connections between tubes and threads are made with a quick-connect adapter. The same applies to the connection between the chamber and the solenoid valves.

\paragraph{Connection between electronic components:} Electronic components, including MFCs, solenoid valves, and the IR camera, have their unique connector pairs in the PCB. We power the module (PCB, RPi, controllers, and sensors) using a 24 \si{\volt}, 20 \si{\ampere} AC/DC converter, connected directly to the PCB.

\begin{figure}[H]
        \centering
        \includegraphics[width=0.8\linewidth]{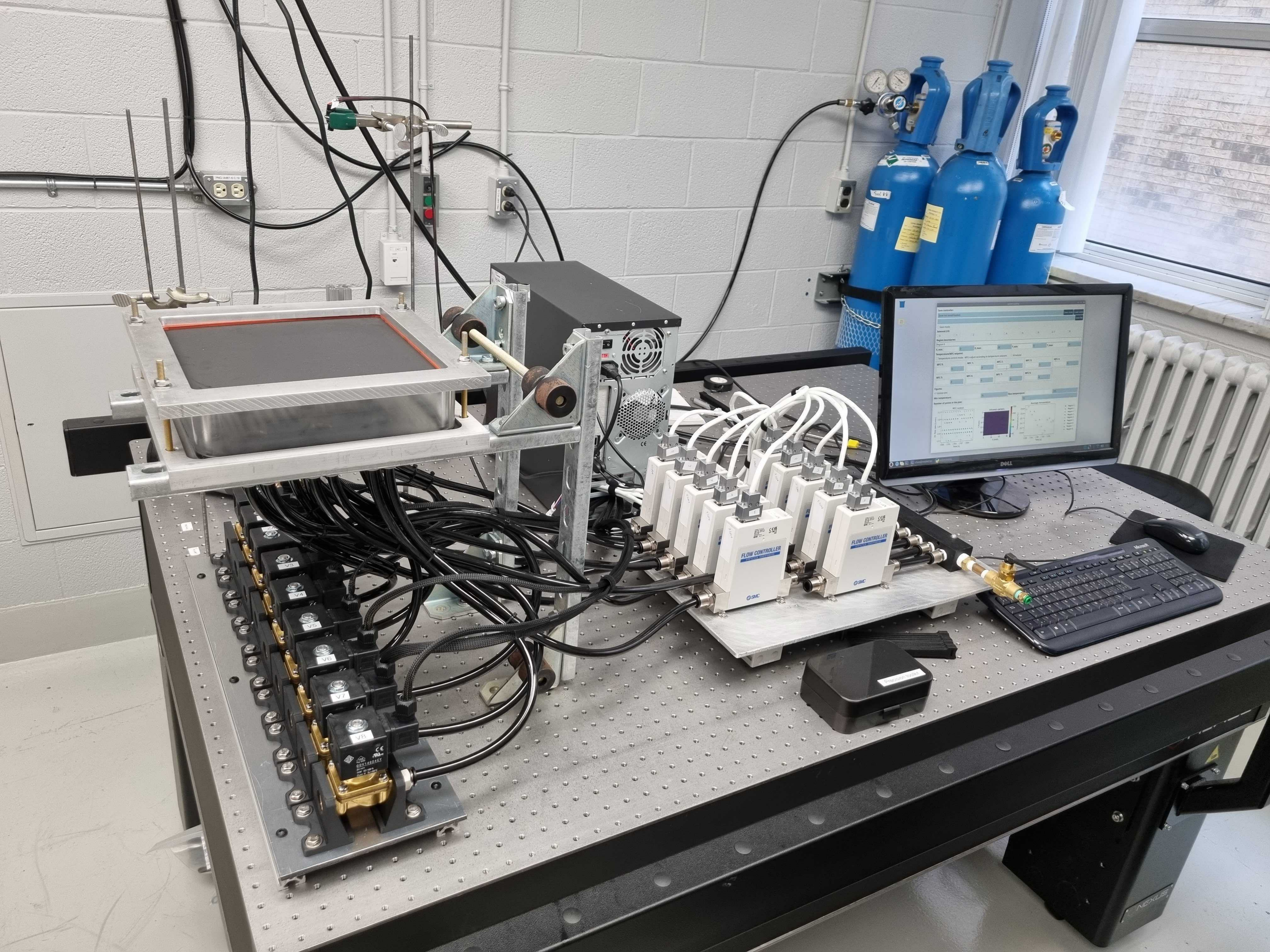}
        \caption{Picture of the setup mounted on the optical table. It includes all parts. Control parts, including the PCB and Raspberry Pi, are inside the computer casing in the image.}
        \label{fig:final}
\end{figure}

The fully assembled system is presented in Fig. \ref{fig:final}.

\section{Operation instructions}\label{sec:operation}

\begin{figure}[H]
        \centering
        \includegraphics[width=0.9\linewidth]{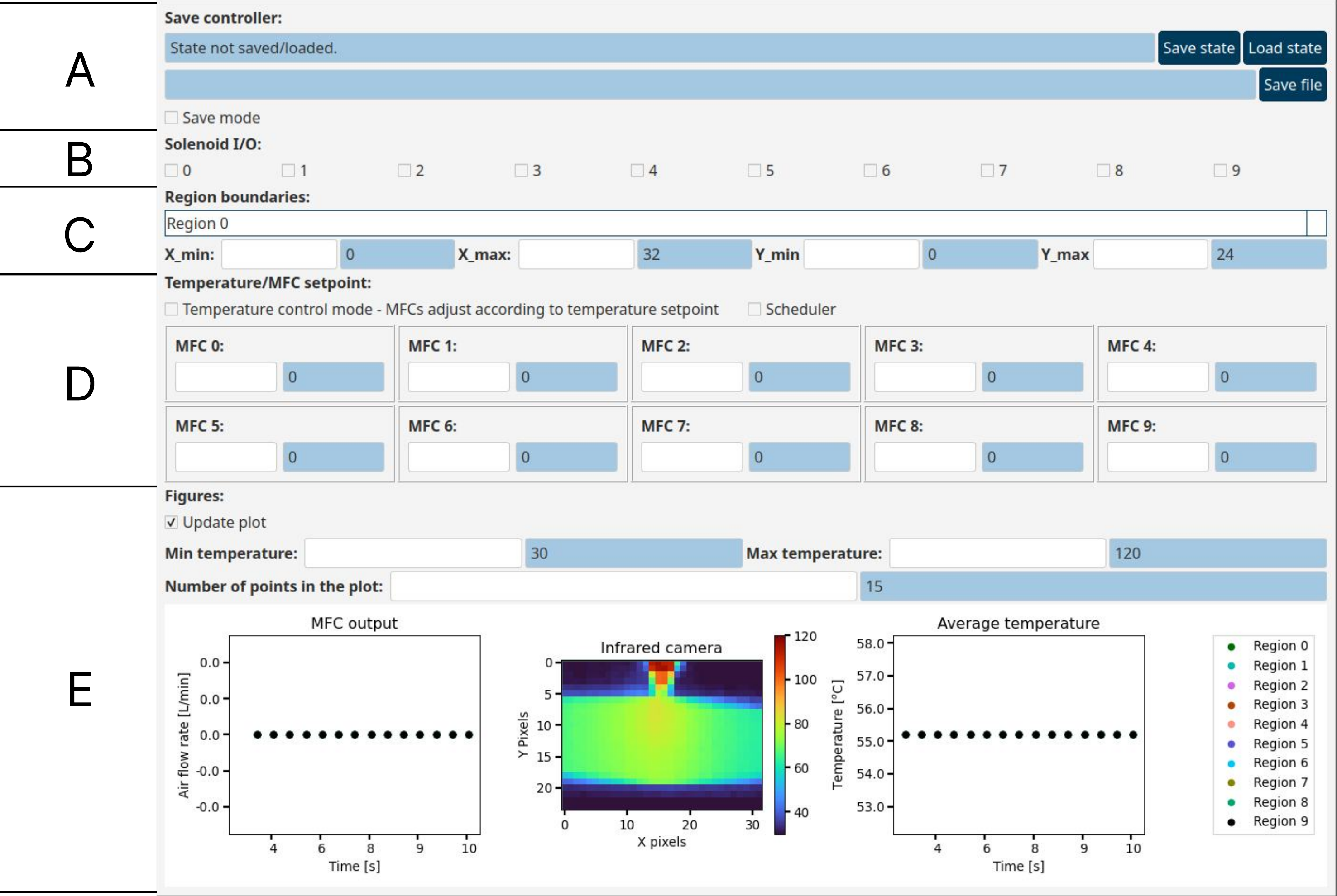}
        \caption{Full Graphical User Interface (GUI).}
        \label{fig:gui}
\end{figure}

We operate the equipment through the GUI application. The application is shown in Figure \ref{fig:gui}. In this section, the operation of the equipment is explained by describing each individual part of the GUI.

\paragraph{Region boundaries:} First, we need to assign a sub-domain to each MFC to establish their respective zones of influence. These zones are rectangular, and typically, their centers coincide with the orifice axis. This is done by inputting the minimum and maximum point coordinates in the x and y axes. By default, when the application starts, all regions cover the entire field of view of the camera, with x going from 0 to 32 and y from 0 to 24.

The number of regions and MFCs is the same. We define the boundary limits of individual sections using the drop down menu indicating ``Region 0''  in Figure \ref{fig:gui} (C).

\paragraph{Solenoid I/O:} The solenoid valves control is done using the checkboxes in Figure \ref{fig:gui} (B). Solenoid valves are mapped to their own checkboxes and respond immediately to changes in checkbox state.

\paragraph{Temperature/MFC setpoint:} The device can take MFC flow rate or target temperature as input. When the ``Temperature control mode'' checkbox is unchecked (default), the flow rate per MFC is controlled directly by the user. In this mode, each MFC has a framed box allowing the user to manually input the flow rate. Figure \ref{fig:gui} (D) illustrates the use of the interface for the operation with two MFCs. The MFC mode allows us to input the flow rate desired per MFC in \si{\liter \per \minute}. This input becomes the new flow rate set point, constantly displayed in the right text box. When we operate the experimental setup illustrated in Figure \ref{fig:cad_iso}, the number of MFCs is 9, and so is the number of framed boxes in this section of the GUI.

\begin{figure}[H]
        \centering
        \includegraphics[width=0.9\linewidth]{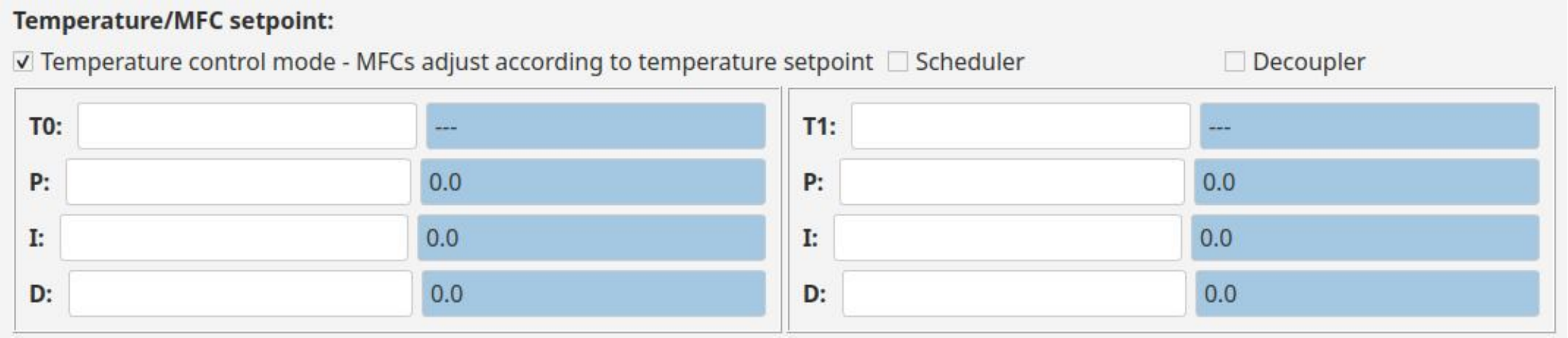}
        \caption{Temperature setpoint and controller gains definition.}
        \label{fig:gui_pid}
\end{figure}

As shown in Figure \ref{fig:gui_pid}, when the ``temperature control mode'' checkbox is checked, the framed boxes change and the temperature control parameters become available. This mode takes the temperature setpoint and manipulates MFC flow rates to achieve it. Similarly to the MFC mode, when a parameter is set in the left text box, it is displayed in the right text box. The temperature control in the example is done by using a Proportional-Integral-Derivative (PID) control algorithm \cite{seborg_process_2017}. The parameters of this controller are the temperature setpoint and the gains of each component. The software gathers the average temperature per region and applies the controller output to the MFC flow rate. The control mode also allows for the addition of a standard decoupler, which can be added by checking the checkbox in Figure \ref{fig:gui_pid}.

\paragraph{Scheduler: } When the ``Scheduler'' option is checked (unchecked by default) the user is prompted to select a scheduler file. A CSV file, with the first column specifying time and subsequent columns specifying flow rate values for each MFC, allows for automated tests. In ''Temperature Control Mode'', the scheduler uses the file to automate setpoint change tests and the specified values correspond to temperature setpoints. The UI displays the start and end values of the time interval, along with either the flow rate of each MFC or the specified temperature setpoints for each region, depending on the selected mode.

\paragraph{Figures: } Allows the user to track the experiment information in real-time. It includes three graphics: air flow rate per MFC as a function of time, temperature heat map, and average temperature per region as a function of time. The regions' placement and boundaries are illustrated as boxes in the heat map, part (E) of Figure \ref{fig:gui}.

The temperature range in the heat map can be modified by changing the value in the ``Min temperature'' and ``Max temperature'' boxes. We can also change the number of previous points in the time series using the ``Number of points in the plot'' setting.

Lastly, we use the ``Update plot'' checkbox (checked, by default) to determine whether the plots are updated or not. By not updating the graphics, we reduce the computational cost of the script execution per iteration of the application, which allows us to reduce the acquisition interval if necessary.

\paragraph{Save controller:} This section enables the selection of the directory of the data and state files. The state files store information such as the region boundaries and PID gains of the controllers for efficient experiment reproduction. In other words, reproducing an experimental condition only implies loading its state file. Saving and loading the files is as easy as clicking the buttons next to the text line, which contains the path to the current state file (if one exists).

``Save file'' function produces two data files. When in ``MFC control mode'', it contains elapsed time since the last initialization of  ``Save mode'' (unchecked by default), the flow rates per MFC, the average temperature per region, and the region border points. When in ``Temperature control mode'' it additionnally contains the temperature setpoints and the proportional, integral and derivative gains per MFC. The second output file records the time and the temperature per pixel within the heat map. 

\section{Experimental demonstration}\label{sec:experiments}

We conducted three experiments to demonstrate the performance of the setup: a bump test to assess the system's response to a step change in mass flow rate, a setpoint tracking experiment to evaluate its ability to regulate the temperature and a disturbance rejection experiment with two PI-controlled inlets. In the latter, a localized heat gun was successively applied above each inlet, showing that the heat load can be relocated and that each PI controller primarily regulates the temperature in the zone directly above its corresponding inlet.

These tests characterize the system’s thermal response and control performance. The tests served as an initial assessment of the setup’s fundamental thermal and fluid dynamic behavior while allowing for future control strategy development. All results presented in this section have been processed using a Savitzky-Golay filter with a window length of 10 and a polynomial order of one to reduce noise while preserving flow rate and temperature variations.

\subsection{Step response characterisation}

\begin{figure}[H]
        \centering
        \includegraphics[width=0.9\linewidth]{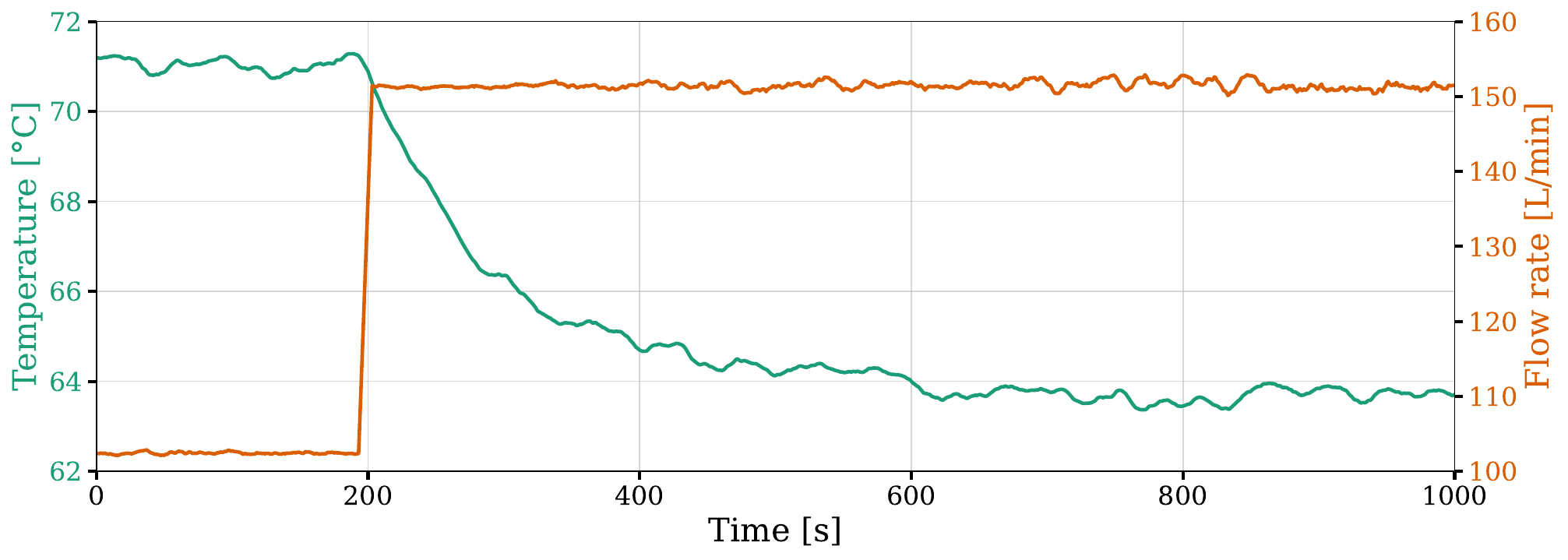}
        \caption{System Response to a Step Change in Flow Rate.}
        \label{fig:bump-test}
\end{figure}

To quantify the system's response to a change in mass flow rate, a bump test was performed with a single mass flow controller (MFC). The flow rate was increased from 100 to 150 \si{\liter \per \minute}, resulting in a temperature drop from 71 to 64 \si{\celsius}, with a response time of 400 \si{\second}. Figure \ref{fig:bump-test} shows both the user-imposed step change in flow rate and the resulting temperature response. A transient phase is observed after the flow rate bump before reaching a new steady-state. 

Performing a series of such bump tests allows for the characterization of the response of the system and provides data for controller tuning.

\subsection{Setpoint Tracking Performance}

A setpoint change experiment was performed to demonstrate the system's ability to regulate the temperature using the implemented PI control algorithm. For this experiment, the following parameters were used: proportional gain = 10.46, integral gain = 0.11. These values were obtained using the direct synthesis tuning method \cite{seborg_process_2017}. The target temperature was decreased from 98 to 75 \si{\celsius}.

\begin{figure}[H]
    \centering
    \includegraphics[width=0.9\linewidth]{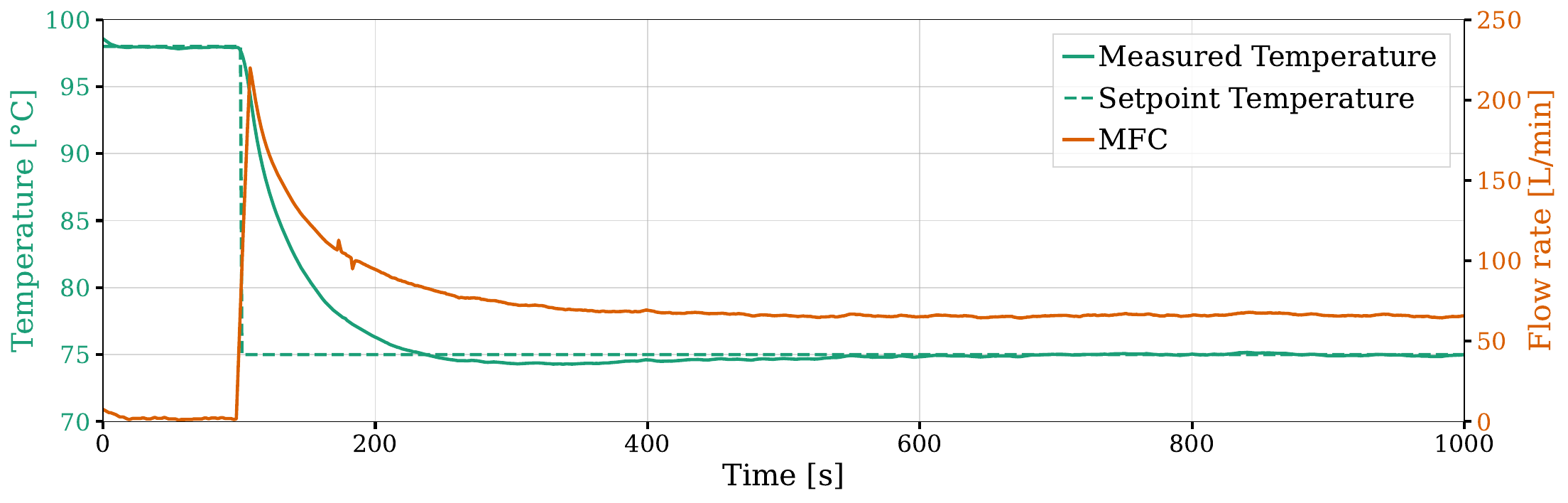}
    \caption{System Response to a Setpoint Change in Temperature.}
    \label{fig:setpoint-tracking}
\end{figure}

Figure \ref{fig:setpoint-tracking} illustrates both the temperature response and the corresponding change in flow rate following the setpoint decrease. After an initial peak in flow, the flow rate stabilizes around 75 \si{\liter \per \minute}, as the system reaches steady-state.

These experiments demonstrate the unit’s ability to adapt to setpoint changes by actuating on the mass flow to ensure precise and efficient thermal management. The results confirm that the hardware can effectively regulate temperature, with response characteristics indicating potential for applications requiring controlled heat extraction over time and space.

\subsection{Disturbance rejection performance with a localized heat load}

\begin{figure}[H]
    \centering
    \includegraphics[width=0.9\linewidth]{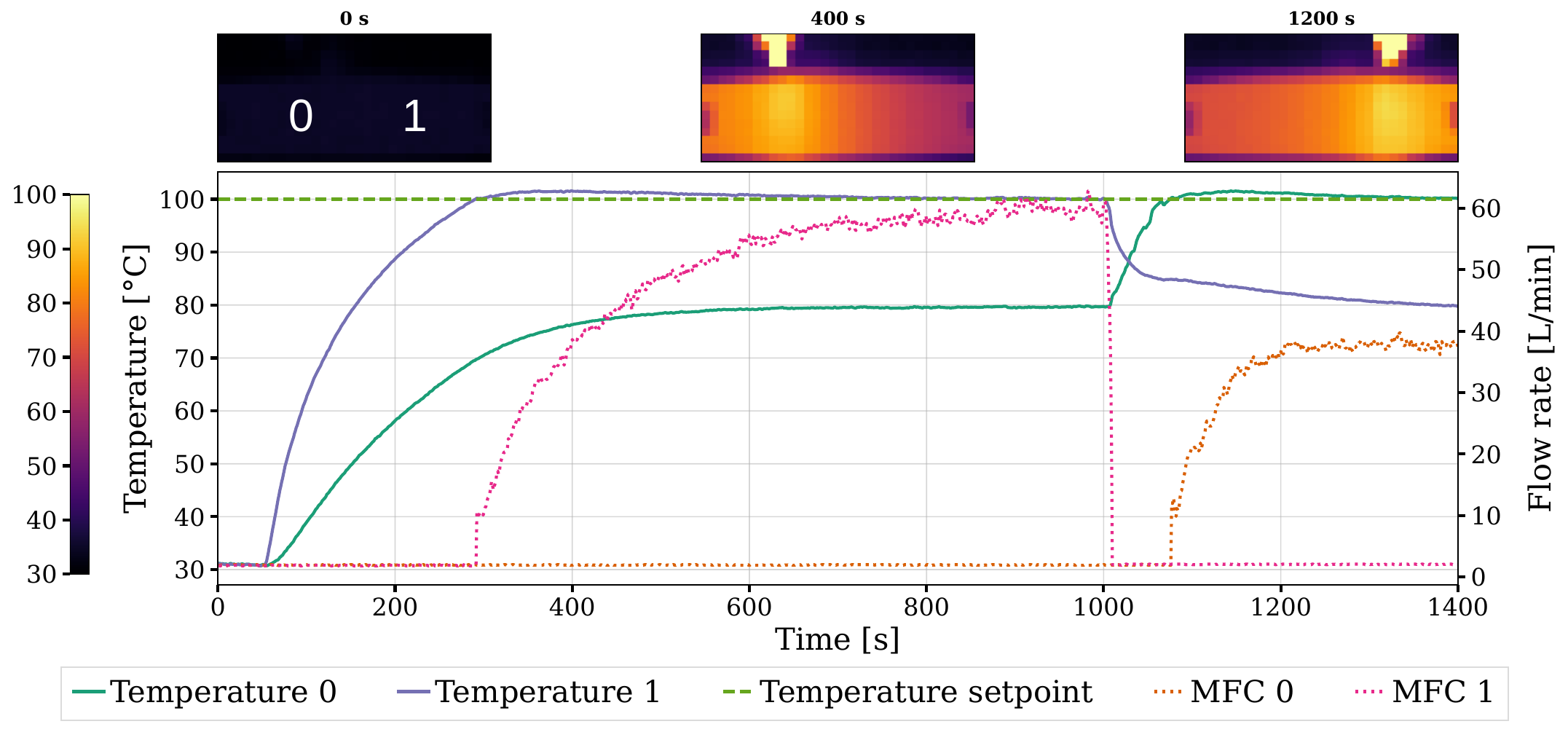}
    \caption{System Response of two PI-controlled inlets to a localized heat-gun disturbance. The heat gun is applied over zone 1, then repositioned to be over zone 2 at 1000 \si{\second}.}
    \label{fig:disturbance-rejection}
\end{figure}

Disturbance rejection was evaluated with two independent PI controllers, each regulating the surface temperature over a 3x3 pixel region centered above its respective jet inlet. The jet arrangement from left to right was: outlet, inlet (MFC 1), closed, inlet (MFC 0), outlet. For this experiment both controllers used the following parameters: proportional gain = 10, integral gain = 0.1, and derivative gain = 0. The temperature setpoint under heat load was \SI{100}{\celsius} and the experiment timeline is shown in Figure \ref{fig:disturbance-rejection}. 

The system was initially undisturbed. At 50 \si{\second}, a heat gun was positioned above region 0, causing the local temperature to rise. MFC 1 responds once the measured temperature in its control region exceeds its setpoint. By 1000 \si{\second}, the system reached a new steady state. The heat gun was then moved from zone 1 to zone 0, resulting in a decrease in the flow rate of MFC 1 as its region cooled, while MFC 0 increased its flow rate to counter the new disturbance. 

This simple experiment demonstrated that the disturbance location can be shifted during operation and that each PI controller acts locally to regulate the temperature directly above its corresponding inlet. Notably, the two zones have different temperatures, never reaching the setpoint simultaneously. However, the device enables further characterization of the system and tuning of the control strategy.

\section{Conclusion}

This work demonstrates a simple, modular, and versatile experimental setup to safely test spatio-temporal thermal management strategies using technology developed by \citet{Lamarre_2023}. The provided files include CAD drawings, a PCB design, and a totally custom and user-friendly software to control the device. We also demonstrate use cases of the proposed design, including system characterization and control performance. The modular nature of the software allows users to leverage existing components to build their own application in order to interface with sub-components of the experimental setup, such as the MFCs and the solenoid valves. The openly available database simplifies extending the application to testing varied control strategies and enables adaptation of the design to similar thermal management devices.

\noindent
\\
\textbf{CRediT author statement}\\

\noindent
{\textbf{Victor Oliveira Ferreira}: Conceptualization, Investigation, Methodology, Software, Validation, Formal analysis, Writing. \textbf{Wiebke Mainville}: Conceptualization, Investigation, Validation, Formal analysis, Writing. \textbf{Vincent Raymond}: Conceptualization, Investigation, Methodology, Resources, Funding acquisition. \textbf{Jean-Michel Lamarre}: Conceptualization, Investigation, Funding acquisition. \textbf{Antoine Hamel}: Conceptualization, Investigation, Methodology, Software. \textbf{Mikael Vaillant}: Conceptualization, Investigation, Formal analysis. \textbf{Moncef Chioua}: Conceptualization, Investigation, Methodology, Formal analysis, Supervision. \textbf{Bruno Blais}: Conceptualization, Investigation, Methodology, Formal analysis, Writing, Supervision, Project administration, Funding acquisition.}\\

\noindent
\textbf{Acknowledgements}\\

This work was funded by the National Research Council of Canada's (NRC) National Program Office (NPO) through the Advanced Manufacturing program (AM-126-2). This research was enabled in part by support provided by Calcul Qu\'ebec (https://www.calculquebec.ca/) and the Digital Research Alliance of Canada (https://alliancecan.ca/en).

\bibliographystyle{unsrtnat}
\bibliography{references}

\begin{appendix}

\begin{landscape}

\section{Bill of materials}\label{sec:bom}

\centering
\begin{table}[htpb!]
\resizebox{0.85\linewidth}{!}{\begin{tabular}{|l|l|l|l|l|l|l|}
\hline
\textbf{Designator} & \textbf{Component} & \textbf{Number} & \textbf{Cost per unit - USD} & \textbf{Total Cost - USD} & \textbf{Source of materials (store name)} & \textbf{Material type} \\ \hline
C74 & Capacitor ALUM 100UF 20\% 35V SMD & 2 & \$                       1.05 & \$                   2.10 & DigiKey Canada & N/A \\ \hline
C75, C76, C77 & Capacitor ALUM 100UF 20\% 50V SMD & 1 & \$                       0.43 & \$                   0.43 & DigiKey Canada & N/A \\ \hline
C56, C57 & Capacitor ALUM 470UF 20\% 50V RADIAL TH & 1 & \$                       1.66 & \$                   1.66 & DigiKey Canada & N/A \\ \hline
C31, C33, C35, C37, C47, C49,   C55, C71, C72 & Capacitor CER 0.1UF 50V X7R 0402 & 3 & \$                       0.05 & \$                   0.16 & DigiKey Canada & N/A \\ \hline
C1, C2, C5, C6, C7, C8, C9,   C10, C12, C13, C16, & Capacitor CER 10000PF 50V X7R 0402 & 1 & \$                       0.00 & \$                   0.00 & DigiKey Canada & N/A \\ \hline
C52, C53, C54, C59, C62 & Capacitor CER 10UF 50V X5R 1206 & 8 & \$                       0.11 & \$                   0.84 & DigiKey Canada & N/A \\ \hline
C3, C4, C11, C14, C15, C22,   C30, C32, C70, C73 & Capacitor CER 1UF 25V X5R 0402 & 26 & \$                       0.07 & \$                   1.82 & DigiKey Canada & N/A \\ \hline
C58, C60, C61, C67, C68, C69,   C78 & Capacitor CER 22UF 25V X5R 1210 & 1 & \$                       0.46 & \$                   0.46 & DigiKey Canada & N/A \\ \hline
C27, C28, C29, C66 & Capacitor CER 4.7UF 25V X5R 0805 & 1 & \$                       0.08 & \$                   0.08 & DigiKey Canada & N/A \\ \hline
C63 & Capacitor CER 4700PF 50V X7R 0603 & 1 & \$                       0.05 & \$                   0.05 & DigiKey Canada & N/A \\ \hline
C64 & Capacitor CER 680PF 50V X7R 0402 & 10 & \$                       0.04 & \$                   0.35 & DigiKey Canada & N/A \\ \hline
2233K19 & Clamping Nut for 1-5/8" Strut Channel & 8 & \$                       3.75 & \$                30.00 & McMaster Carr & Plastic \\ \hline
{\color[HTML]{333333} 8560K914} & clear Scratch- and UV-Resistant Cast Acrylic Sheet, 12" x   12" x 2" & 1 & \$                   147.68 & \$              147.68 & McMaster Carr & Rubber \\ \hline
J2, J3, J8, J9, J16, J17, J18,   J19, J20, J21 & Click-mate 1.5 DRVT SMT AU0.1 ETP & 1 & \$                       2.87 & \$                   2.87 & DigiKey Canada & N/A \\ \hline
 & Connector 24-28AWG CRIMP TIN & 3 & \$                       0.08 & \$                   0.23 & DigiKey Canada & N/A \\ \hline
{\color[HTML]{333333} } & Connector HEADER R/A 4POS 4.2MM & 3 & \$                       1.73 & \$                   5.19 & DigiKey Canada & N/A \\ \hline
J4, J5, J6, J7, J10, J11, J12,   J13, J14, J15 & Connector HEADER VERT 2POS & 2 & \$                       1.16 & \$                   2.32 & DigiKey Canada & N/A \\ \hline
 & Connector Plug HSG 10POS 1.50MM & 2 & \$                       0.42 & \$                   0.83 & DigiKey Canada & N/A \\ \hline
 & Connector Plug HSG 4POS 1.50MM & 1 & \$                       0.41 & \$                   0.41 & DigiKey Canada & N/A \\ \hline
J1, J3, J4 & Connector RCPT HSG 2POS 3.00MM & 1 & \$                       0.51 & \$                   0.51 & DigiKey Canada & N/A \\ \hline
 & Connector RCPT HSG 4POS 4.20MM & 2 & \$                       0.55 & \$                   1.10 & DigiKey Canada & N/A \\ \hline
J22 & Connector RECPT R/A SMC 4 POS & 1 & \$                       1.60 & \$                   1.60 & DigiKey Canada & N/A \\ \hline
 & Connector Socket 16AWG CRIMP TIN & 2 & \$                       0.21 & \$                   0.42 & DigiKey Canada & N/A \\ \hline
 & Connector Socket 20-24AWG CRIMP TIN & 1 & \$                       0.08 & \$                   0.08 & DigiKey Canada & N/A \\ \hline
U16 & DC/DC Converter 3.3-16.5V 100W & 2 & \$                     35.59 & \$                71.18 & DigiKey Canada & N/A \\ \hline
D22, D23, D24, D25, D27, D28,   D29, D30, D32, D33 & Diode GEN PURP 50V 1A SMA & 1 & \$                       0.26 & \$                   0.26 & DigiKey Canada & N/A \\ \hline
D35 & Diode SCHOTTKY 30V 5A SMC & 1 & \$                       0.56 & \$                   0.56 & DigiKey Canada & N/A \\ \hline
D34 & Diode ZENER 12V 500MW SOD123 & 1 & \$                       0.09 & \$                   0.09 & DigiKey Canada & N/A \\ \hline
PFCQ531-04-A1C-S & Flow Controller for Air, 9-300 lpm & 10 & \$               1,108.21 & \$        11,082.10 & SMC & N/A \\ \hline
F1, F2 & Fuse BRD MT 375MA 125VAC 63VDC & 10 & \$                       0.38 & \$                   3.75 & DigiKey Canada & N/A \\ \hline
F3 & Fuse BRD MT 5A 125VAC 63VDC 1206 & 1 & \$                       0.38 & \$                   0.38 & DigiKey Canada & N/A \\ \hline
1388K454 & Ground Low-Carbon Steel Sheet, 10" x 10" x 1/8" & 1 & \$                     57.32 & \$                57.32 & McMaster Carr & Plastic \\ \hline
3433K55 & Heat Gun, Pistol-Style, & 1 & \$                   212.30 & \$              212.30 & McMaster Carr & N/A \\ \hline
3313N774 & High-Strength Steel Threaded Rod, Zinc Yellow-Chromate Plated,   1/2"-13 Thread Size, 1 Foot Long & 4 & \$                     24.11 & \$                96.44 & McMaster Carr & Metal \\ \hline
U2, U4 & Integrated Circuit Analog-Digital Converter (ADC), 12BIT SAR   16WQFN & 4 & \$                       4.09 & \$                16.36 & DigiKey Canada & N/A \\ \hline
U12 & Integrated Circuit CURRENT MONITOR 5\% 24VQFN & 10 & \$                       4.81 & \$                48.10 & DigiKey Canada & N/A \\ \hline
U7, U8 & Integrated Circuit Digital-Analog Converter (DAC), 12BIT V-OUT   24 VQFN & 2 & \$                     17.27 & \$                34.54 & DigiKey Canada & N/A \\ \hline
U14 & Integrated Circuit linear regulator 5.5V 300MA SOT23-5 & 10 & \$                       0.56 & \$                   5.60 & DigiKey Canada & N/A \\ \hline
U1, U3, U5 & Integrated Circuit OPAMP GP 4 CIRCUIT 14TSSOP & 5 & \$                       2.62 & \$                13.10 & DigiKey Canada & N/A \\ \hline
U15 & Integrated Circuit REG LINEAR 3.3V 500MA DPAK & 30 & \$                       2.54 & \$                76.20 & DigiKey Canada & N/A \\ \hline
U6, U17 & Integrated Circuit XLTR VL BIDIR 8-VSSOP & 1 & \$                       1.30 & \$                   1.30 & DigiKey Canada & N/A \\ \hline
D36 & LED green clear SMD & 1 & \$                       0.23 & \$                   0.23 & DigiKey Canada & N/A \\ \hline
D21, D26, D31 & LED Red clear SMD & 11 & \$                       0.23 & \$                   2.53 & DigiKey Canada & N/A \\ \hline
Q2 & MOSFET N-CH 30V 2.2A SUPERSOT3 & 1 & \$                       0.96 & \$                   0.96 & DigiKey Canada & N/A \\ \hline
Q1 & MOSFET P-CH 40V 90A TO252-3 & 1 & \$                       4.00 & \$                   4.00 & DigiKey Canada & N/A \\ \hline
8983K361 & Multipurpose 304 Stainless Steel Sheet & 1 & \$                     67.38 & \$                67.38 & McMaster Carr & Metal \\ \hline
8975K87 & Multipurpose 6061 Aluminum sheet, 1/2" & 1 & \$                     11.51 & \$                11.51 & McMaster Carr & Metal \\ \hline
9246K13 & Multipurpose 6061 Aluminum sheet, 3/4" & 1 & \$                     43.89 & \$                43.89 & McMaster Carr & Metal \\ \hline
9452K448 & Oil-Resistant Buna-N O-Ring, 1/4 Fractional Width, Dash Number   454 & 1 & \$                     11.42 & \$                11.42 & McMaster Carr & Metal \\ \hline
5225K516 & Push-To-Connect Tube Fitting For Air, Straight Adapter, 12 Mm   Tube Od x 1/2 Bspp & 18 & \$                     26.60 & \$              478.80 & McMaster Carr & Plastic \\ \hline
5225K718 & Push-To-Connect Tube Fitting For Air, Straight Adapter, For 12   Mm Tube Od x 3/8 & 13 & \$                     18.49 & \$              240.37 & McMaster Carr & Plastic \\ \hline
SC1113 & Raspberry Pi 5, 16 GB, 2.4 GHz 64-bit Quad Core ARM Processor,   Dual-Band 802.11ac Wi-Fi, Bluetooth 5.0 & 1 & \$                   174.53 & \$              174.53 & DigiKey Canada & N/A \\ \hline
R29, R30, R37, R38, R57, R61,   R62, R63 & Resistor 0 Ohm Jumper 1/16W 0402 & 3 & \$                       0.01 & \$                   0.03 & DigiKey Canada & N/A \\ \hline
R1, R3, R5, R8, R10, R12, R14,   R16, R19, R21, R5 & Resistor 100 Ohm 1\% 1/16W 0402 & 1 & \$                       0.01 & \$                   0.01 & DigiKey Canada & N/A \\ \hline
R18 & Resistor 100K Ohm 1\% 1/16W 0402 & 1 & \$                       0.01 & \$                   0.01 & DigiKey Canada & N/A \\ \hline
R48 & Resistor 100K Ohm 1\% 1/16W 0402 & 1 & \$                       0.04 & \$                   0.04 & DigiKey Canada & N/A \\ \hline
R51 & Resistor 1M Ohm 1\% 1/16W 0402 & 1 & \$                       0.01 & \$                   0.01 & DigiKey Canada & N/A \\ \hline
R52 & Resistor 330 Ohm 1\% 1/16W 0402 & 7 & \$                       0.01 & \$                   0.08 & DigiKey Canada & N/A \\ \hline
R7 & Resistor 33K Ohm 1\% 1/16W 0402 & 10 & \$                       0.01 & \$                   0.11 & DigiKey Canada & N/A \\ \hline
R23, R24, R27, R28, R31, R32,   R33, R34, R35, R36 & Resistor 470 Ohm 1\% 1/16W 0402 & 2 & \$                       0.01 & \$                   0.02 & DigiKey Canada & N/A \\ \hline
R53 & Resistor 56K Ohm 1\% 1/16W 0402 & 1 & \$                       0.01 & \$                   0.01 & DigiKey Canada & N/A \\ \hline
R39, R40, R41 & Resistor 6.8K Ohm 1\% 1/16W 0402 & 5 & \$                       0.01 & \$                   0.06 & DigiKey Canada & N/A \\ \hline
R46 & Resistor SMD 0 Ohm Jumper 1/2W 1206 & 3 & \$                       0.71 & \$                   2.13 & DigiKey Canada & N/A \\ \hline
R2, R4, R6, R9, R11, R13, R15,   R17, R20, R22, R2 & Resistor SMD 10K Ohm 1\% 1/16W 0402 & 1 & \$                       0.02 & \$                   0.02 & DigiKey Canada & N/A \\ \hline
R45, R49 & Resistor SMD 20K Ohm 1\% 1/5W 0402 & 20 & \$                       0.09 & \$                   1.74 & DigiKey Canada & N/A \\ \hline
R56 & Resistor SMD 7.87K Ohm 1\% 1/10W 0402 & 9 & \$                       0.04 & \$                   0.39 & DigiKey Canada & N/A \\ \hline
98150A170 & Slide-Adjust Push-Button Nut, Hex, 1/2"-13 Thread Size & 4 & \$                     17.14 & \$                68.56 & McMaster Carr & Plastic \\ \hline
98150A770 & Slide-Adjust Push-Button Nut, Knurled-Head, 1/2"-13 Thread Size & 8 & \$                     17.14 & \$              137.12 & McMaster Carr & Metal \\ \hline
1023N279 & Straight-Flow Rectangular Manifold, Anodized Aluminum,   Standard, 10 Outlets, 1/2 Npt x 3/8 Npt & 1 & \$                     63.19 & \$                63.19 & McMaster Carr & Metal \\ \hline
{\color[HTML]{333333} 33125T601} & Strut Channel Bracket, 90 Degree Angle, Zinc-Plated Steel,   4" x 4" Long & 4 & \$                     15.24 & \$                60.96 & McMaster Carr & Metal \\ \hline
{\color[HTML]{333333} 3259T31} & Strut Channel Nut, Zinc-Plated Steel, 1/2"-13 Thread Size, 3/8"   High & 4 & \$                       8.77 & \$                35.08 & McMaster Carr & Metal \\ \hline
3310T517 & Strut Channel, Low-Profile,   Slotted Hole, Galvanized Steel & 4 & \$                       9.38 & \$                37.52 & McMaster Carr & Metal \\ \hline
2545T81 & Super-Conductive 101 Copper, Softened Temper Sheet, 12" x   12", 1/8" Thick & 1 & \$                   120.92 & \$              120.92 & McMaster Carr & Metal \\ \hline
D37 & TVS Diode 24VWM 38.9VC SMA & 1 & \$                       0.47 & \$                   0.47 & DigiKey Canada & N/A \\ \hline
D1, D2, D3, D4, D5, D6, D7,   D8, D9, D10, D11, D1 & TVS Diode 5VWM 9.2VC DO219AB & 1 & \$                       0.22 & \$                   0.22 & DigiKey Canada & N/A \\ \hline
D38 & TVS Diode 80VWM SOT363 & 10 & \$                       0.17 & \$                   1.66 & DigiKey Canada & N/A \\ \hline
51305K429 & Universal-Thread Push-To-Connect Tube Fitting For Air And   Water, Wye, 12 Mm Tube Od x 1/2 Male Pipe & 9 & \$                     11.53 & \$              103.77 & McMaster Carr & Metal \\ \hline
N/A & Viper VP4000 Mini 500GB M.2 2230 PCIe Gen4 x4 SSD - Solid   State Drive - VP4000M500GM23 & 1 & \$                     60.99 & \$                60.99 & DigiKey Canada & N/A \\ \hline
\end{tabular}}
\label{tab:BOM}
\end{table}

\end{landscape}

\end{appendix}

\end{document}